\documentclass[10pt,twoside,onecolumn]{article}
\usepackage[]{latexsym,amsmath,amssymb}
\usepackage[]{epsfig}
\newcommand{\be}{\begin{eqnarray}}
\newcommand{\ee}{\end{eqnarray}}
\newcommand{\ud}{\mathrm{d}}

\title{\bf Asymptotic Safety, Singularities, and Gravitational Collapse}
\author{Roberto Casadio$^{ab}$\thanks{casadio@bo.infn.it}$\ $,
Stephen D.H. Hsu$^c$\thanks{hsu@uoregon.edu}$\ $
and Behrouz Mirza$^d$\thanks{b.mirza@cc.iut.ac.ir}
\\
\\
$^a$Dipartimento di Fisica, Universit\`a di Bologna,
via~Irnerio~46, 40126~Bologna, Italy
\\
$^b$
I.N.F.N.,
Sezione di Bologna, via~Irnerio~46, 40126~Bologna, Italy
\\
$^c$Institute of Theoretical Science, University of Oregon,
Eugene, OR 97403-5203
\\
$^d$Department of Physics, Isfahan University of Technology,
 Isfahan  84156-83111, Iran}
\begin{document}
\maketitle
\begin{abstract}
Asymptotic safety (an ultraviolet fixed point with finite-dimensional critical surface) offers
the possibility that a predictive theory of quantum gravity can be obtained from the quantization
of classical general relativity.
However, it is unclear what becomes of the singularities of classical general relativity, which, it
is hoped, might be resolved by quantum effects.
We study dust collapse with a running gravitational coupling and find that a future singularity
can be avoided if the coupling becomes exactly zero at some finite energy scale.
The singularity can also be avoided (pushed off to infinite proper time) if the coupling approaches
zero sufficiently rapidly at high energies.
However, the evolution deduced from  perturbation theory still implies a singularity at finite proper time.
\end{abstract}
%
%
%
%
\section{Introduction}
\setcounter{equation}{0}
It is an appealing possibility that classical general relativity can be directly quantized,
leading to a predictive and well-defined theory of quantum gravity.
Asymptotically free models such as QCD are potentially valid to arbitrarily short distances,
with quarks and gluons remaining the appropriate degrees of freedom.
Similarly, if gravity exhibits asymptotic safety~\cite{weinberg}, an ultraviolet fixed point with
finite-dimensional critical surface, gravitons can remain fundamental at all energy scales,
and low-energy physics is determined by only a finite number of parameters.
(The corresponding literature is very extensive; for a partial selection see
Refs.~\cite{AS,Weinberg-inflation}).
However, in such scenarios it remains to understand how, or whether, the singularities of
classical general relativity~\cite{hawking} are resolved by quantum effects.
It would be disturbing if a truly fundamental theory of gravitation contained singularities.
\par
To investigate this question, we consider one of the simplest physical situations which
leads to a future singularity: the collapse of a pressureless ball of dust~\cite{OS}.
This system collapses under its own weight into a point-like singularity
in a finite amount of proper time, and is the historical prototype of stellar collapse
to form a black hole.
Of course, one expects matter pressure to be relevant in a realistic case,
however, in classical general relativity, one can always assume a star is sufficiently
massive to overcome the effect of pressure and produce a black hole
(if the mass is approximately twice the Chandrasekhar limit~\cite{chandra}).
\par
Let us point out some of the limits of our approach.
Our calculations are {\it inspired\/} by asymptotic safety, but we cannot derive the form
of the modified Einstein equations used below.
Also, we cannot treat the case in which higher dimension operators become important
to the evolution of the dust ball, as semi-classical equations of motion resulting from
such operators would presumably be higher than second order in derivatives.
However, such behavior would usually be associated with a strongly coupled fixed point.
It is hard to see how a singularity could be avoided in such a scenario unless, for instance,
at least some gravitational interactions became strongly {\it repulsive\/} (rather than attractive)
at the fixed point. It is also hard to see how one could reliably establish the existence of
a strongly coupled fixed point without truly non-perturbative methods.
\section{Dust collapse}
\setcounter{equation}{0}
\label{dust}
We shall here consider the spatially flat Tolman-Lemaitre-Oppenheimer-Snyder
model of a dust ball which collapses under its own gravitational
pull~\cite{OS}.
The outer ($R>R_s$) metric is the Schwarzschild line element
\be
\ud s^2=-\left(1-\frac{2\,M_s}{R}\right)\ud t^2
+\left(1-\frac{2\,M_s}{R}\right)^{-1}\ud R^2
+R^2\,\ud\Omega^2
\ ,
\ee
where $M_s$ is the total ADM mass (with units of length)
of the ball with areal radius $R=R_s$.
The interior geometry is given by
\be
\ud s^2=-\ud\tau^2+(R')^2\,\ud r^2+R^2\,\ud\Omega^2
\ ,
\ee
where $0\le r\le r_s$, $\tau$ is the proper time of observers
comoving with the dust along lines of constant $r$,
the areal radius $R=R(\tau,r)$, $R'=\partial_r R$ and
$\dot R=\partial_\tau R$.
The relevant equations are given by the Einstein equation
\be
\frac{(\dot R^2\,R)'}{R^2\,R'}=8\,\pi\,G_{\rm N}\,\rho
\ ,
\label{Gtt}
\ee
where $\rho=\rho(\tau,r)$ is the dust energy density,
and the Bianchi identity
\be
\dot\rho+\frac{\partial_\tau(R^3)'}{(R^3)'}\,\rho=0
\ .
\label{bianchi}
\ee
The Bianchi identity implies the conservation of the Tolman mass function
\be
m(r)=\frac{4\,\pi}{3}\,\int_0^r\rho(\tau,x)\,[R^3(\tau,x)]'\,dx
\ ,
\ee
for all $0<r\le r_s$.
Note that the corresponding ``ADM length'' is simply given by
\be
M(r)=G_{\rm N}\,m(r)
\ ,
\ee
and this is the quantity that naturally enters the equations of motion for
dust (see below).
\par
In particular, we can write
\be
\label{ra}
R=r\,a(\tau)
\ ,
\ee
for $0\le r\le r_s$ and $\tau\ge \tau_0\equiv 0$ (the arbitrary initial time of the collapse).
The standard junction conditions at the surface of the ball $r=r_s$ then yield the trajectory
$R_s(\tau)=r_s\,a(\tau)$ in the outer Schwarzschild space-time, with constant ADM length
$M_s=M(r_s)$.
Eq.~\eqref{Gtt} can then be used to express the density as a function of $a=a(\tau)$ and
$G_{\rm N}$, namely
\be
\rho=\rho(\tau)=\frac{3}{8\,\pi\,G_{\rm N}}   \left( \frac{\dot a}{a} \right)^2
\ ,
\ee
so that
\be
M(r)=\frac{1}{2}\,r^3\,a\,\dot a^2
\ ,
\ee
and dust trajectories at constant $0<r\le r_s$ automatically satisfy the equation of motion
\be
\dot R^2=\frac{2\,M}{R}
\ ,
\ee
which, for $G_{\rm N}$ constant, leads to the well-known expression
\be
a(\tau)=\left(1-\frac{\tau}{\tau_s}\right)^{2/3}
\ ,
\label{ac}
\ee
where we set $a(\tau_0=0)=1$ without loss of generality
and $\tau_s>0$ is the proper time at which the entire ball becomes singular.
\section{Running coupling and singularities}
\setcounter{equation}{0}
\label{AS}
We next assume that Eqs.~\eqref{Gtt} and \eqref{bianchi} still hold for
a Newton constant that depends on an energy scale sensitive to local conditions
({\em i.e.}, at a particular region of space-time).
The natural choice for our simple system is thus the energy density, that is
$G_{\rm N}\to G(\rho)$, with $G(\rho)=G_{\rm N}$ for sufficiently low density.
We first proceed to analyze the behavior of $G(\rho)$ that follows from the
equations of motion.
We will later introduce an explicit renormalization scale $\mu$, along with
a proper definition of ultra-violet (UV) fixed point, and discuss its relation with $\rho$
more carefully.  
It is however important to note from the onset that, if Eq.~\eqref{bianchi} does not hold,
the ADM length $M_s$ of the ball is not conserved and the outer space-time
may not be Schwarzschild.
This would be tantamount to a large scale (IR) modification to general relativity,
with potentially observable consequences, {\em e.g.}, in solar system tests~\cite{will}.
\par
The modified Einstein equation~\eqref{Gtt} is simply given by
\be
\label{eeq}
\left( \frac{\dot{a}}{a} \right)^2 = \frac{ 8 \,\pi}{3}\, \rho\, G( \rho )
\ .
\ee
Substituting Eq.~\eqref{ra} into the Bianchi identity~\eqref{bianchi} implies
\be
\label{d1}
\frac{\dot{\rho}}{\rho} = - 3 \, \frac{\dot{a}}{a}
\ ,
\ee
or
\be
\label{d2}
\frac{\rho (\tau)} {\rho (\tau_0)}=  \left[\frac{a(\tau)}{a(\tau_0)} \right]^{-3}
\ .
\ee
Note this relation is independent of any assumptions about the scale dependence of $G$.
\par
From Eq.~\eqref{eeq} and Eq.~\eqref{d2}, we see that if $G( \rho )$ becomes zero at  some
critical energy scale $\rho_*$, the scale factor $a(\tau)$ and density $\rho(\tau) = \rho_*$
also become constant, thereby avoiding a singularity.
If $G( \rho \geq \rho_*) = 0$, this behavior is non-analytic (it implies discontinuous derivatives
of $G$ with respect to the energy scale).
However, since in most models the UV fixed point (see below) is approached smoothly
as the energy scale goes to infinity, we shall only consider the case of $G(\rho)$
analytic in $\rho$.
\par
Combining Eq.~\eqref{eeq} and Eq.~\eqref{d1}, we obtain
\be
\label{integral}
\int_{\rho(\tau_0=0)}^{\rho (\tau)} \frac{\ud\rho}{\sqrt{24\, \pi\, \rho^3\, G( \rho ) }} = \tau
\ ,
\ee
in which we set $\tau_0=0$ again without loss of generality.
Let us assume that $G (\rho )$ is always positive.
If the integral in Eq.~\eqref{integral} is convergent as $\rho(\tau)$ (the upper limit of integration)
goes to infinity, then there must exist some finite $\tau_*$ at which
$\rho(\tau \to \tau_*) \to \infty$ and  $a(\tau \to \tau_*)\sim \rho^{-1/3} \to 0$ [see Eq.~\eqref{d2}]
-- {\em i.e.}, a singularity at finite proper time.
This would be the case for constant $G$ (ordinary classical gravity), a non-zero fixed point
value of the coupling [$G(\rho \to \infty) \neq 0$)], and for $G( \rho )$
which approaches zero sufficiently slowly at high energy.
Alternatively, if $G (\rho )$ falls to zero rapidly enough, the integral above may be divergent,
allowing for the possibility that the singularity is pushed off to infinite proper time:
$\rho(\tau \to \infty) \to \infty$ and $a(\tau \to \infty) \to 0$.
For example, if $G ( \rho ) \sim 1 / \rho$, we obtain $\rho(\tau)$ and $a(\tau)$ which
are asymptotically exponential in $\tau$.
\par
Using perturbation theory ({\em i.e.}, ordinary Feynman graphs)
one obtains the following leading-order evolution for the
gravitational coupling as a function of the scale $\mu$:
\be
\label{pert}
\frac{1}{G(\mu)} = \frac{1}{G(\mu=0)} + c\, \mu^2
\ ,
\ee
where $c > 0$ for models in which the coupling becomes weaker
at higher energies.
A UV fixed point is then defined in terms of the dimensionless Newton
constant $g(\mu)=G(\mu)\,\mu^2$ as the limiting value $g(\mu\to\infty)= g_*<\infty$,
which requires $G(\mu)\to 0$ (as previously mentioned).
\par
The form~\eqref{pert} for the evolution of $G(\mu)$ is also expected
from simple dimensional analysis.
In Ref.~\cite{calmet}, it is noted that using a generally covariant
heat kernel regularization results in positive contributions to
$c$ from spin-1 degrees of freedom ({\em e.g.}, gauge bosons) and
negative contributions from spin-0 and spin-1/2 modes.
Thus, a gravity model with sufficiently large number of photon-like
matter fields is perturbatively under control, that is
\be
G ( \mu \to \infty) \to 0
\ee
with
\be
\label{af2}
g(\mu)=G( \mu )\,\mu^2 \ll 1
\ . 
\ee
The latter condition in fact implies that even the shortest distance quantum gravity effects
are small, and perturbation theory applies at all times.
This is because the dimensionless parameter in the gravitational loop expansion
is given by the high energy cutoff squared times the effective coupling at that scale.
Strictly speaking, asymptotic freedom obtains for $g(\mu\to\infty)=0$.
\par
To pursue our analysis further, we must determine more carefully the relationship
between the renormalization scale $\mu$ and the density $\rho$.
One appealing choice, advocated by Weinberg in his analysis of inflation in
asymptotically safe gravity~\cite{Weinberg-inflation}, is to take the renormalization
group mass scale $\mu$ to be 
\be
\label{rg}
\mu \sim \left[G (\mu) \, \rho\right]^{1/2}
\ ,
\ee
which has the appearance of the inverse of a ``gravitational length'' related to the energy
density $\rho$ and is equivalent to taking $\mu$ to be the inverse of the timescale over which
the scale factor $a( \tau )$ changes.
If we were to take $G (\mu)$ to be constant in Eq.~\eqref{rg}, we would obtain
the asymptotic behavior
\be
\label{asymptotic2}
G(\rho)\sim 1 / \rho
\ee
at high density, which is just sufficient to push the singularity off to infinite proper time
[see Eq.~\eqref{integral}].
However, if a self-consistently defined $G( \mu )$ (which decreases with increasing
$\mu$ or $\rho$) is used in Eq.~\eqref{rg}, the behavior is insufficient to prevent
a singularity.
\par
For definiteness, let us assume the asymptotic form
\be
G(\mu)\sim \mu^{-\alpha}
\ ,
\label{asyG}
\ee
with $\alpha>0$ and constant and, in particular, $\alpha\ge 2$ for asymptotic
safety.
With this choice we exclude the case of $G(\mu)$ which increases for increasing $\mu$
and do not further consider the case $\alpha=0$ corresponding
to $G=G_{\rm N}$ and constant.
Eq.~\eqref{rg} then implies
\be
\rho\sim\mu^{2+\alpha}
\quad
{\rm and}
\quad
G(\rho)\sim \rho^{-\frac{\alpha}{2+\alpha}}
\ .
\ee
Plugging the above expressions into Eq.~\eqref{integral} yields the following
asymptotic solution:
\be
\begin{array}{l}
\rho(\tau\to\tau_*)\sim \left(\tau_*-\tau\right)^{-(2+\alpha)}
\\
\\
a(\tau\to\tau_*)\sim \left(\tau_*-\tau\right)^{(2+\alpha)/3}
\\
\\
\mu(\tau\to\tau_*)\sim \left(\tau_*-\tau\right)^{-1}
\\
\\
G(\tau\to\tau_*)\sim \left(\tau_*-\tau\right)^{\alpha}
\ ,
\end{array}
\label{AS2_1}
\ee
which contains a singularity (of infinite density and vanishing radius) at finite proper
time $\tau=\tau_*$.
Note that for any choice of $\alpha$ the renormalization scale $\mu$ is of order the
inverse proper timescale, which seems physically reasonable.
The condition $\mu^2\, G( \mu ) \ll 1$ [from Eq.~\eqref{af2}], necessary for short distance
perturbative control of the theory, implies $\alpha \ge 2$ (and asymptotic freedom
requires $\alpha > 2$), so models in which perturbative
analysis is self-consistent are quite far from avoiding a singularity. 
\section{Conclusions}
\setcounter{equation}{0}
We studied how the simplest model of gravitational collapse in classical general
relativity would be modified by a gravitational coupling that runs with the energy scale.
Our aim was to determine under which conditions the final singularity could be avoided
in an asymptotically safe scenario. 
Our findings seem to imply that avoiding the final singularity requires stronger deviations
from general relativity than those obtained in the literature this far, unless a more
realistic model of gravitational collapse changes the equations that govern $G$ as a
function of the local energy scale significantly.
\section*{Acknowledgements}
R.C.~is supported by INFN grant BO11. S.H.~thanks D.~Reeb for discussions and is supported
by the Department of Energy under grant DE-FG02-96ER40969.
B.M.~ thanks the Institute of Theoretical Science for its hospitality.

\begin{thebibliography}{99}
%
%
\bibitem{weinberg}
S.~Weinberg, ``Ultraviolet divergences in quantum theories of gravitation,''
in~{\em General relativity: an Einstein centenary survey},
edited by S.~Hawking and W.~Israel,
Cambridge University Press (Cambridge, 1979).
%
\bibitem{AS}
C.~Wetterich,
  Phys.\ Lett.\  B {\bf 301}, 90 (1993);
M.~Reuter,
  Phys.\ Rev.\  D {\bf 57}, 971 (1998)
  [arXiv:hep-th/9605030];
W.~Souma,
  Prog.\ Theor.\ Phys.\  {\bf 102}, 181 (1999)
  [arXiv:hep-th/9907027];
M.~Niedermaier,
  Class.\ Quant.\ Grav.\  {\bf 24}, R171 (2007)
  [arXiv:gr-qc/0610018];
M.~Niedermaier and M.~Reuter,
  Living Rev.\ Rel.\  {\bf 9}, 5 (2006);
R.~Percacci and D.~Perini,
  Phys.\ Rev.\  D {\bf 68}, 044018 (2003)
  [arXiv:hep-th/0304222];
R.~Percacci and D.~Perini,
  Phys.\ Rev.\  D {\bf 67}, 081503 (2003)
  [arXiv:hep-th/0207033];
S.~P.~Robinson and F.~Wilczek,
  Phys.\ Rev.\ Lett.\  {\bf 96}, 231601 (2006)
  [arXiv:hep-th/0509050];
R.~Percacci,
  [arXiv:hep-th/0709.3851];
K.~Falls, D.~F.~Litim and A. Raghuraman,
  [arXiv:hep-th/1002.0260];
Y.~Cai and D.~A.~Esson,
  [arXiv:hep-th/1007.1317];
S.~Weinberg,
   Phys.\ Rev.\  D {\bf 81}, 083535 (2010)
  [arXiv:hep-th/0911.3165].
%

\bibitem{Weinberg-inflation}
S.~Weinberg,
  Phys.\ Rev.\  D {\bf 81}, 083535 (2010)
  [arXiv:0911.3165 [hep-th]].


\bibitem{hawking}
  S.~W.~Hawking and G.~F.~R.~Ellis,
  {\em The Large scale structure of space-time},
Cambridge University Press (Cambridge, 1973).
%
\bibitem{OS}
  J.~R.~Oppenheimer and H.~Snyder,
  Phys.\ Rev.\  {\bf 56}, 455 (1939);
%
\bibitem{chandra}
  S.~Chandrasekhar,
  Astrophys.\ J.\  {\bf 74}, 81 (1931).
%
\bibitem{will}
  C.~M.~Will,
  Living Rev.\ Rel.\  {\bf 9}, 3 (2005)
  [arXiv:gr-qc/0510072].
%
\bibitem{calmet}
X.~Calmet, S.~D.~H.~Hsu and D.~Reeb,
  Phys.\ Rev.\ Lett.\  {\bf 101}, 171802 (2008)
  [arXiv:0805.0145 [hep-ph]].
%
\end{thebibliography}
\end{document}